\documentclass[twocolumn,preprintnumbers,amsmath,amssymb,pre]{revtex4}
\usepackage{epsfig,graphicx}

\usepackage{dcolumn,textcomp}
\usepackage{bm}

\begin{document}

\title{Complete breakdown of the Debye model of rotational relaxation near
the isotropic-nematic phase boundary: Effects of intermolecular correlations
in orientational dynamics}

\author{Prasanth P. Jose, Dwaipayan Chakrabarti and  Biman Bagchi}
\email{bbagchi@sscu.iisc.ernet.in}
\homepage{http://sscu.iisc.ernet.in/prg/faculty/biman_bagchi.htm}
\affiliation{Solid state and Structural Chemistry Unit, Indian Institute of Science,
Bangalore 560012, Karnataka, India.}

\begin{abstract}
The Debye-Stokes-Einstein (DSE) model of rotational diffusion predicts 
that the rotational correlation times $\tau_{l}$ vary as $[l(l+1)]^{-1}$,
where $l$ is the rank of the orientational correlation
function (given in terms of the Legendre polynomial of rank $l$). One
often finds significant deviation from this prediction, in either direction.
In supercooled molecular liquids where the ratio $\tau_{1}/\tau_{2}$ 
falls considerably below three (the Debye limit), one usually invokes a jump 
diffusion model to explain the approach of the ratio $\tau_{1}/\tau_{2}$ to 
unity. Here we show in a computer simulation study of a standard model system 
for thermotropic liquid crystals that this ratio becomes 
{\it much less than unity} 
as the isotropic-nematic phase boundary is approached from the isotropic side. 
Simultaneously, the ratio $\tau_2/\eta$ (where $\eta$ is the shear viscosity of the liquid)
becomes {\it much larger} than hydrodynamic value near the I-N transition.
We have also analyzed the break down of the Debye model of rotational diffusion 
in ratios of higher order rotational correlation times. We show that the
break down of the DSE model is due to the growth of orientational pair 
correlation and provide a mode coupling theory analysis to explain the
results.
\end{abstract}
\maketitle

\section{Introduction}

The rotational diffusion model of Debye \cite{debye,bap} was proposed
originally to explain dielectric relaxation of polar molecules and
to relate the observed relaxation time to viscosity by the use of the
Stokes-Einstein relation. The Debye-Stokes-Einstein (DSE) model is 
primarily devised to model the Brownian motion in orientational degrees 
of freedom. The DSE model provides the following amazingly simple expression 
for the decay of the $l$-th rank orientational correlation function $C_{l}^{s}(t)$
\begin{equation}
C_{l}^{s}(t)=exp(-t/\tau_{l}),\label{eq:clt}\end{equation}
with\begin{equation}
\tau_{l}^{-1}=l(l+1)k_{B}T/\zeta_{R},\label{eq:taul}\end{equation}
where
\begin{equation}
\zeta_{R}=6\eta V\label{eq:fric}\end{equation}
In the above expression of rotational friction, $V$ is the volume
of the molecule in question, $\eta$ is the viscosity of the liquid,
$k_{B}$ is Boltzmann constant and $T$ is the temperature. The 
above expressions 
predict the ratio of the first and second-rank rotational correlation
times, $\tau_{1}/\tau_{2}$, to be equal to three. The higher rank rotational
correlation times also vary following the $[l(l+1]^{-1}$ dependence. 
In the past, different aspects of DSE 
relationship have been tested
namely, (a) the shape dependence 
\cite{alb-dbye,shape:debye,zwanzig:debye1}, (b) the viscosity 
dependence, and (c) the rank dependence of the orientational
relaxation,  in simple as well as complex liquids.

In many cases, the ratio $\tau_{1}/\tau_{2}$ is found to fall below
three. In particular, in supercooled liquids, the ratio is known to approach unity at low 
temperatures and related issues (like translation-rotation decoupling) that 
have given rise to considerable amount of discussion
\cite{still1:ng,still2:ng,
scool:kiv1,scool:kam,scool:ngai1,scool:edig1,scool:sill2,
scool:lap2,scool:lap3,scool:lap4,
scool:fdebye,scool:corez}. One 
often invokes the breakdown of Debye diffusion
model, which requires small angle Brownian rotational motion for its
validity, due to the emergence of large scale hopping involving large angular 
jumps. The basic idea is that a large jump leads
to the decay of $C_{l}^{s}(t)$ of all rank $l$ at the same time,
so that all of them have similar correlation times corresponding to the average
waiting time for this large jump to occur.

The case of $\tau_{1}/\tau_{2}$ deserves particular attention because
it is the oft discussed one. The sensitivity of these two correlation
times (and hence the respective correlation functions) to intermolecular
interactions is different. When the rotating molecule has dipolar
interactions with the surrounding molecules, one expects the first
rank rotation to be more affected than the second rank one. This is
a manifestation of intermolecular correlation which is reflected in
the orientational pair correlation functions. Thus, if we consider
dipolar spheres (such Stockmayer liquid), we should expect the ratio
$\tau_{1}/\tau_{2}$ to be larger than three. On the other hand, if
the intermolecular interaction has up-down symmetry (as in the case
of molecules with ellipsoid of revolution), the reverse could be true.
Thus, one may need to include the role of intermolecular correlation
to generalize the DSE model appropriately. 

One ideal candidate to test the above argument regarding the role
of equilibrium pair correlations is a system of model mesogens undergoing 
the isotropic-nematic (I-N) phase transition. The 
isotropic phase is both positionally and orientationally disordered while the 
nematic phase is still positionally disordered 
but orientationally ordered. Earlier experimental and simulations 
studies \cite{bb:fay1,bb:fay2,fay2,fay3,
bb:ppj1,bb:ppj3,bb:dwc} have demonstrated that existence of the orientational 
relaxation of nematogens near I-N transition have similarities with that 
observed in the supercooled liquids. The orientational order parameter is 
defined by $S=\left<P_{2}(cos(\theta))\right>$, where $\theta$ is the angle of 
the orientation of the molecular axis with the director. In the isotropic phase, 
$S=0$ for infinitely large systems while it is non-zero (around 0.5) in the 
nematic phase. The important 
point is that orientational correlation undergoes rapid increase as the 
I-N phase boundary is approached from the isotropic side. This large growth in 
correlation would provide a testing ground of the rotational diffusion model 
and the role of intermolecular correlations.

Different molecular models have been used to test the DSE in molecular
simulations. In a molecular dynamics simulation study using the Gay-Berne 
intermolecular potential \cite{deMug2}, de Miguel {\it et al.} 
found the ratio $\tau_{1}/\tau_{2}$ to be always less than
three in the isotropic phase. They found that the departure from the
Debye limit was pronounced when density was reduced or temperature was
increased; this deviation is a manifestation of the inertial decay which
gives a ratio $[l+1/l]$ for $\tau_{l}/\tau_{l+1}$. 
Vasanthi {\it et al.} in extensive molecular dynamics simulations 
using the same model have studied the aspect ratio dependence of the DSE 
relationship \cite{bb:vas2}. Recently Jose and Bagchi have studied the 
breakdown of the DSE relationship near the I-N phase transition in a system of 
the Gay-Berne ellipsoids of revolution \cite{bb:ppj2}. They have shown that the 
relation between the rotational friction and viscosity breaks down near the 
I-N phase transition.

The motivation of the present work comes partly from recent reports
of the observed similarity in the orientational relaxation between
supercooled liquids and liquid crystals.\cite{bb:fay1,bb:fay2,fay2,fay3,
bb:ppj1,bb:ppj3,bb:dwc}
The deviation of the rank dependence from the Debye model (Eq.2) has
been often discussed in the context of supercooled liquids. As discussed
earlier, this is attributed to the existence of large angular jumps
at low temperatures.
In this work, we investigated the rank dependence of the rotational
diffusion near the isotropic-nematic phase boundary. We find that the 
ratio $\tau_{1}/\tau_{2}$ becomes much less than unity 
as the isotropic-nematic phase boundary is approached from the isotropic side. 
Simultaneously, the ratio $\tau_2/\eta$ (where $\eta$ is the shear viscosity of the liquid)
becomes much larger than hydrodynamic value near the I-N transition.
We have also analyzed the breakdown of the Debye model of rotational diffusion 
in ratios of higher order rotational correlation times. Theoretical 
analysis shows that the
breakdown of the DSE model can be attributed to the growth of orientational pair correlation. We provide a mode coupling theory analysis to explain the
results. Thus, the present analysis seems to suggest that one need not 
always invoke large scale jump diffusion 
to explain the decrease of the ratio $\tau_{1}/\tau_{2}$ from the Debye limit. This view raises some interesting questions which we address
in the Conclusion.

In the next section, we describe the system and simulation details. 
 Results of our molecular dynamics simulation study of a system of ellipsoids of 
revolution with an aspect ratio equal to three along an isotherm and 
an isochore across the I-N phase transition is presented in the section
\ref{sec:res}. These results show
that the ratio $\tau_1/\tau_2$ can become much less than unity. This 
section also includes results for higher rank orientational time 
correlation functions (OTCF). In section \ref{sec:ts}, we present a 
theoretical analysis which can explain theoretical
aspects of these results. Section \ref{sec:cr} provides a summary of 
our results and concluding remarks.

\section{System and simulation details\label{sec:sas}}

Here we consider a system of molecules with axial symmetry interacting with
the Gay-Berne (GB) pair potential that has served as a standard model in the 
simulation studies of thermotropic liquid crystals. In the GB pair  
potential \cite{berne1,berne2}, each molecule is assumed to be an ellipsoid 
of revolution having a single-site representation in terms of the position 
${\bf r}_{i}$ of its center of mass and a unit vector ${\bf e}_{i}$ along its 
principal axis of symmetry. The GB interaction between molecules $i$ and $j$ 
is given by
\begin{widetext}
\begin{equation}
U_{ij}^{GB}({\bf r}_{ij},{\bf e}_{i},{\bf e}_{j})  = 
4\epsilon({\bf \hat r}_{ij},{\bf e}_{i},{\bf e}_{j})(\rho_{ij}^{-12} - 
\rho_{ij}^{-6})
\end{equation}
where 
\begin{equation}
\rho_{ij} = \frac{r_{ij} - \sigma({\bf \hat r}_{ij},{\bf e}_{i},{\bf e}_{j})
+ \sigma_{0}}{\sigma_{0}}.
\end{equation}
Here $\sigma_{0}$ defines the cross-sectional diameter, $r_{ij}$ is the 
distance between the centers of mass of molecules $i$ and $j$, and  
${\bf \hat r}_{ij} = {\bf r}_{ij} / r_{ij} $ is a unit vector along the 
intermolecular separation vector ${\bf r}_{ij}$. The molecular shape 
parameter $\sigma$ and the energy parameter $\epsilon$ both depend on the
unit vectors ${\bf e}_{i}$ and ${\bf e}_{j}$ as well as on 
${\bf \hat r}_{ij}$ as given by the following set of equations:
\begin{equation}
\sigma({\bf \hat r}_{ij},{\bf e}_{i},{\bf e}_{j}) = \sigma_{0}\left[1 - 
\frac{\chi}{2} \left\{\frac{({\bf e}_{i}\cdot{\bf \hat r}_{ij} + 
{\bf e}_{j}\cdot{\bf \hat r}_{ij} )^{2}}
{1 + \chi({\bf e}_{i}\cdot{\bf e}_{j})} + 
\frac{({\bf e}_{i}\cdot{\bf \hat r}_{ij} - 
{\bf e}_{j}\cdot{\bf \hat r}_{ij})^{2}}{1 - \chi({\bf e}_{i} \cdot
{\bf e}_{j})}\right\}\right]^{-1/2}
\end{equation}
with $\chi = (\kappa^{2} - 1) / (\kappa^{2} + 1)$ and
\begin{equation}
\epsilon({\bf \hat r}_{ij},{\bf e}_{i},{\bf e}_{j}) = \epsilon_{0}
[\epsilon_{1}({\bf e}_{i},{\bf e}_{j})]^{\nu}
[\epsilon_{2}({\bf \hat r}_{ij},{\bf e}_{i},{\bf e}_{j})]^{\mu}    
\end{equation}
where the exponents $\mu$ and $\nu$ are adjustable, and
\begin{equation}
\epsilon_{1}({\bf e}_{i},{\bf e}_{j}) = 
[1 - \chi^{2}({\bf e}_{i}\cdot{\bf e}_{j})^{2}]^{-1/2}
\end{equation}
and 
\begin{equation}
\epsilon_{2}({\bf \hat r}_{ij},{\bf e}_{i},{\bf e}_{j}) = 1 - 
\frac{\chi ^{\prime}}{2} \left[\frac{({\bf e}_{i}\cdot{\bf \hat r}_{ij} + 
{\bf e}_{j}\cdot{\bf \hat r}_{ij} )^{2}}
{1 + \chi^{\prime}({\bf e}_{i}\cdot{\bf e}_{j})} + 
\frac{({\bf e}_{i}\cdot{\bf \hat r}_{ij} - 
{\bf e}_{j}\cdot{\bf \hat r}_{ij})^{2}}{1 - \chi ^{\prime}({\bf e}_{i} \cdot
{\bf e}_{j})}\right].
\end{equation}
\end{widetext}
with $\chi^{\prime} = (\kappa^{\prime ~ 1/\mu} - 1) / 
(\kappa^{\prime ~ 1/\mu} + 1)$. Here 
$\kappa = \sigma_{ee}/\sigma_{ss}$ is the aspect ratio of the molecule with 
$\sigma_{ee}$ denoting the molecular length along the major axis and 
$\sigma_{ss} = \sigma_{0}$, and 
$\kappa^{\prime} = \epsilon_{ss}/\epsilon_{ee}$, where $\epsilon_{ss}$ and 
$\epsilon_{ee}$ are the depth of the minima of potential for a pair of molecules 
aligned parallel in a side-by-side configuration and end-to-end 
configuration, respectively. It follows that the GB pair potential defines a
family of potential models, each member of which is characterized by a set 
of four parameters ($\kappa, \kappa^{\prime}, \mu$, $\nu$). In the 
present work, we employ the Gay-Berne pair potential with the 
original and most studied parametrization $(3,5,2,1)$ \cite{deMug3}.   

All quantities are given in reduced units defined in terms of the Gay-Berne 
potential parameters $\epsilon_{0}$ and $\sigma_{0}$: length in units of
$\sigma_{0}$, temperature in units of $\epsilon_{0}/k_{B}$, 
 and time in units of $(m\sigma_{0}^{2}/\epsilon_{0})^{1/2}$,
$m$ being the mass of the ellipsoids of revolution. We set the mass as well as 
the moment of inertia of the ellipsoids equal to unity. The simulation is run 
in a microcanonical ensemble with the system in a cubic box with periodic 
boundary conditions. Further details of the simulation can be found elsewhere
\cite{bb:ppj1}.

The system of Gay-Berne ellipsoids of revolution with aspect ratio $3$ is studied
separately along an isotherm with density variation and along an isochore with 
temperature variation across the I-N transition. As compared to the density 
driven transition, the temperature driven I-N transition in the present system
is known to be rather diffuse \cite{deMug3}. Next we present the 
results of our study.

\section{\label{sec:res}Results}

\subsection{Density variation along an isotherm}

\begin{figure}
\includegraphics[width=8.3cm,keepaspectratio]{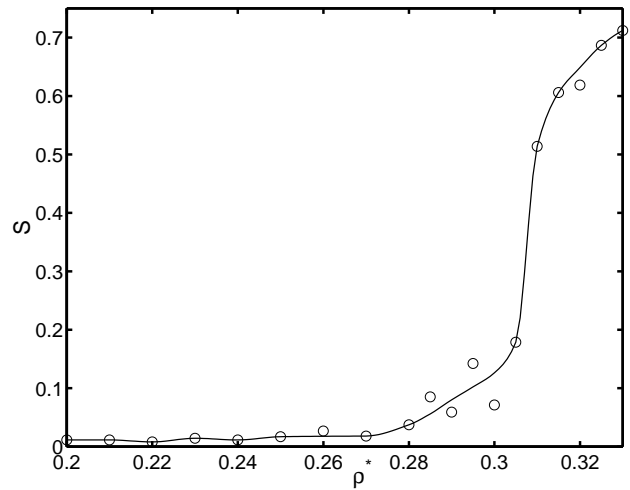}

\caption{\label{cap:op} The evolution of the orientational order parameter 
with density along the isotherm at temperature $T^{\star} = 1$}.
\end{figure}

In figure \ref{cap:op}, we show the orientational order parameter variation 
with density as a system of $576$ Gay-Berne ellipsoids of revolution transits 
across the I-N transition along an isotherm at temperature $T^{\star} = 1$. The phase transition is found 
to occur over a range of density between $0.305$ and $0.315$. %
\begin{figure}
\includegraphics[ width=8.3cm]{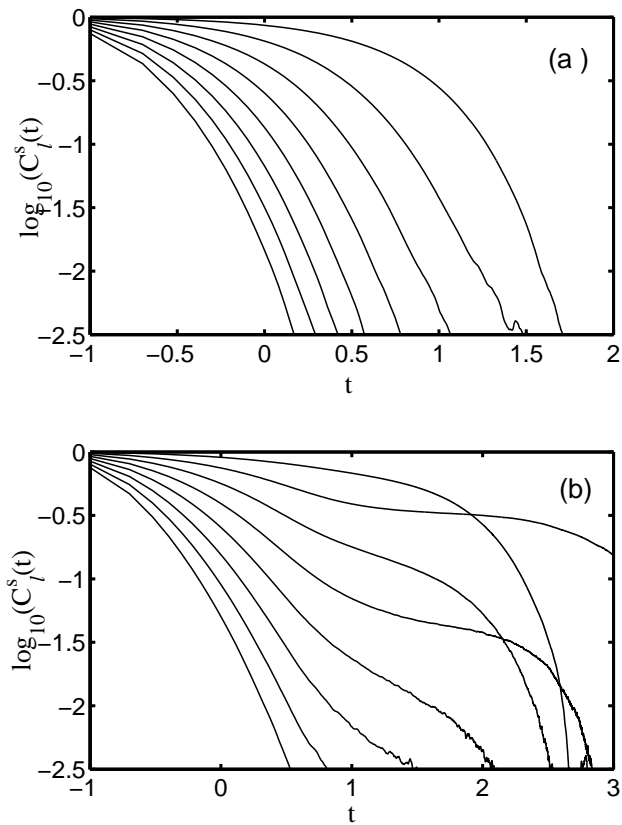}
\caption{\label{cap:csn_de} The time evolution of the single-particle 
orientational time correlation functions, whose rank range from 1 to 8, 
shown in a log-log plot at two densities corresponding to 
(a) $\rho^{*}$= 0.285; (b) $\rho^{*}$=0.315. The curves are arranged
in the decreasing order of ranks from the left to the right in each plot.}
\end{figure}
In figures \ref{cap:csn_de}(a) and \ref{cap:csn_de}(b), we show
the decay of the single-particle orientational correlation function
for the first eight ranks in the isotropic phase and near the I-N phase 
boundary, respectively. The $l$th rank single-particle orientational time 
correlation function (OTCF) is defined as
\begin{equation}
C_{l}^{s}(t)=\frac{\sum_{i}P_{l}(\hat{e}_{i}(0).\hat{e_{i}}(t))}
{\sum_{i}P_{l}(\hat{e}_{i}(0).\hat{e_{i}}(0))},\label{eq:cls}
\end{equation}
where $\hat{e}_{i}(0)$ is the unit vector along the symmetry axis of the 
$i$th ellipsoid of revolution. The DSE model is found to hold good for all ranks
of the single-particle orientational time correlation functions shown in 
Fig. \ref{cap:csn_de}(a) in the isotropic phase. As the rank of the
correlation function increases, the relaxation time decreases. 

Near the I-N phase boundary, as shown in Fig.\ref{cap:csn_de}, the relaxation 
of the single-particle OTCFs slows down considerably for all ranks. However, 
$C_{l}^{s}(t)$ gets affected differently for different $l$ values. The even and the odd $l$-th $C_{l}^{s}(t)$ behave differently with the appearance
of a pronounced plateau in $C_{2}^{s}(t)$ and $C_{4}^{s}(t)$. 
Interestingly, a similar decay behavior has been observed in supercooled 
liquids \cite{scool:kam,scool:lap2}. 

Figure 2(b) shows that although the initial decay of 
$C_{2}^s(t)$ and $C_{4}^s(t)$ is
faster than that of $C_{1}^s(t)$ and $C_{3}^s(t)$, respectively, the decay
becomes slower at longer times. The even l-th correlation functions both
develop a rather long and distinct plateau. The decay of $C_{4}^s(t)$ 
is particularly revealing because it shows all the four phases of decay
 -- the initial Gaussian, followed by the exponential, then the crossover to the plateau and the final exponential decay.

\begin{figure}
\includegraphics[width=8.3cm]{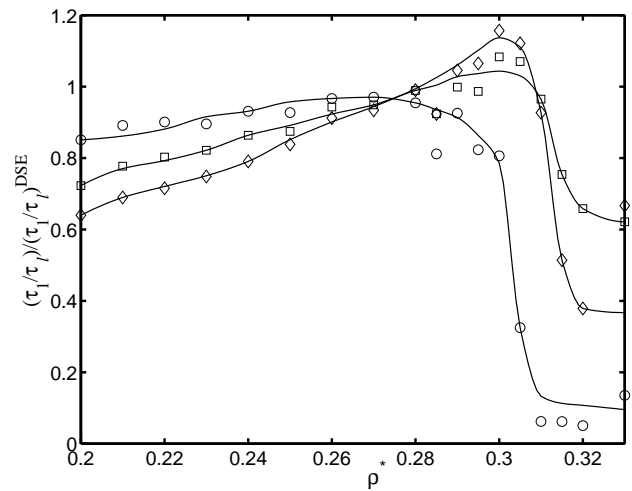}
\caption{\label{cap:ta1} The density variation of the ratios between the 
first-rank orientational correlation time and the second, third, and fourth 
rank orientational correlation times across the I-N transition. The circles 
represent the data for  $\tau_{1}/\tau_{2}$, the diamonds for 
$\tau_{1}/\tau_{3}$ and the triangles for $\tau_{1}/\tau_{4}$. The ratios are 
scaled by the corresponding Debye predictions.}
\end{figure}
In figure \ref{cap:ta1}, we show the evolution of the ratios
$\tau_{1}/\tau_{2}$, $\tau_{1}/\tau_{3}$, and $\tau_{1}/\tau_{4}$ as increase
in density drives the system across the I-N transition. Note that away
from the I-N phase boundary in the isotropic phase, the ratios remain close to 
what are predicted by the Debye rotational diffusion model. However, as the I-N 
phase boundary is approached from the isotropic side, it is evident that the 
Debye rotational diffusion model breaks down completely.
\begin{figure}
\includegraphics[width=8.3cm]{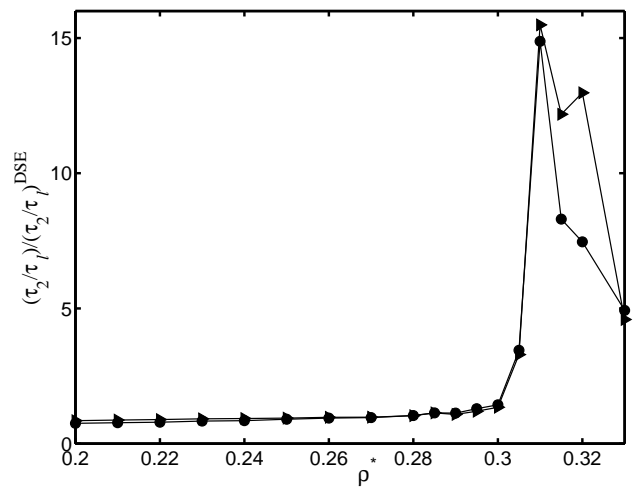}
\caption{\label{cap:tau2} The ratios between the second-rank orientational 
correlation time and the higher rank orientational correlation times across the
I-N transition. The triangles represent the data for $\tau_{2}/\tau_{3}$ and 
the circles for $\tau_{2}/\tau_{4}$. The ratios are scaled by the corresponding
Debye predictions.}
\end{figure}

In figure \ref{cap:tau2}, we show the ratios between the second rank 
orientational correlation time and the higher rank ($l$ = 3,4) orientational 
correlation times as a function of density across the I-N transition. Note that
the I-N transition affects the second rank OTCF most due to the up-down 
symmetry of the molecular model studied in the present work. Figure 
\ref{cap:tau2} shows a cusp like behavior which is well-known
in the study of equilibrium critical phenomena of finite sized systems.
Its appearance in dynamics suggests the existence of large scale fluctuations
in the orientational order parameter \cite{deg:5}. Since the second rank
orientational correlation function is associated with the optical response
of the system which undergoes dramatic increase near the I-N transition 
\cite{deg:5,sch}, the most affected orientational memory function is the 
second-order one.  

Microscopically this phenomenon may be understood qualitatively in terms of the
molecular field theory of Maier and Saupe \cite{deg:5,sch,rmp1}, where the 
molecule is confined in an effective field created by its neighbors. This 
effective potential is given by the expression
\begin{equation}
u_i=-\frac{A}{V^2}\frac{1}{2}S(3cos^2(\theta_i)-1)
\end{equation}
where $A$ is constant independent of the temperature, volume, and pressure, 
$V$ is the molecular volume and $\theta_i$ is the angle between molecular
axis with a preferred axis. This effective potential grows
as the order parameter increases. Note that a $\pi$ rotation of the molecular
axis relaxes $C_1^s(t)$ but not $C_2^s(t)$. This effect is manifested in 
the higher order orientational correlation functions also. However, a 
random orientation of the smaller angle less than $\pi/2$ is only required 
for the relaxation of the $C_l^s(t)$ with $l\geq$ 3,4, .. etc. Hence the 
slow down of relaxation at these ranks appears as $S$ becomes significantly 
large. In section IV, we present a quantitative theory to describe these 
effects.
\begin{figure}
\includegraphics[width=8.3cm]{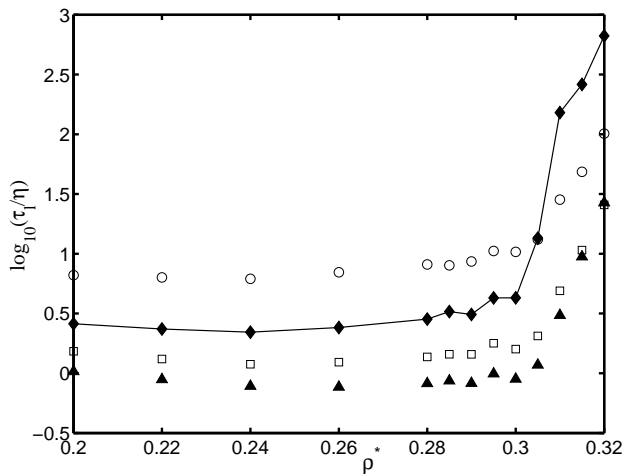}
\caption{\label{cap:taubyeta} The semi-log plot of the ratio $\tau_l/\eta$
versus density for different $l$ values. The circles show $\tau_l/\eta$ for 
$l$ = 1, the diamonds for $l$ = 2, the squares for $l$ = 3, and the triangles 
for $l$ = 4. The filled symbols show the ratios for even values of $l$.}
\end{figure}

Another important aspect of the DSE which has been a subject of intense study
in the literature of supercooled liquids is the viscosity ($\eta$)
dependence of the 
time constant ($\tau_l$) of orientational correlation function.
In figure \ref{cap:taubyeta}, we present the semi-log plot of this
ratio against density. The ratio remains a constant in the isotropic phase of 
the liquid crystals, in agreement with hydrodynamic prediction. 
This is also in accordance with the earlier simulation \cite{bb:ppj2}
which shows the ratio between the rotational friction and the viscosity also 
shows similar behavior. 

Note that unlike in supercooled liquids, the growth in the 
viscosity of the nematogens near the I-N transition is not rapid \cite{bb:ppj2}.
Therefore, the ratio between $\tau_l$ and $\eta$ deviates from the DSE 
prediction only due to the growth of $\tau_l$. It is also evident that the 
violation of the DSE is found to be different for the odd and the even values 
of $l$. In figure \ref{cap:taubyeta}, the ratio between $\tau_l$ and $\eta$ for 
even $l$ values grows more dramatically than that for odd $l$ values. 

\subsection{Temperature variation along an isochore}

\begin{figure}
\includegraphics[width=7.0cm,angle=270]{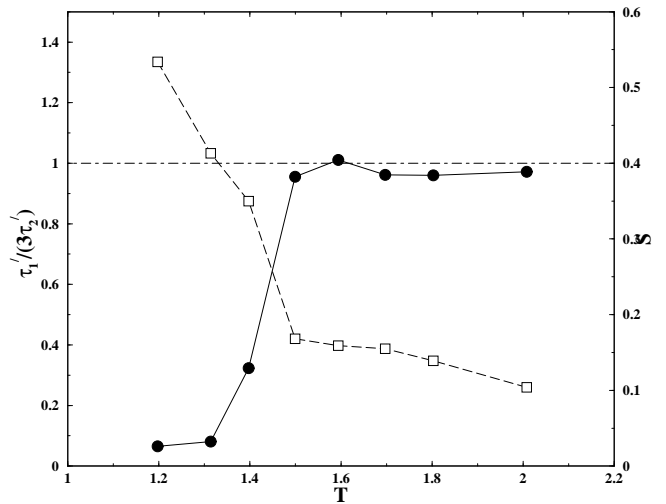}
\caption{\label{cap:tmp-tau2b1-op} The evolution of the ratio between the 
first-rank and the second-rank orientational correlation times with temperature 
across the I-N transition (circles). The inclusion of the scaling factor 
ensures that the ratio is equal to unity in the Debye limit as shown by the 
dot-dashed line. On a different scale shown on the right is the 
orientational order parameter variation with temperature (squares).}
\end{figure}
\begin{figure}
\includegraphics[ width=7.0cm, angle=270]{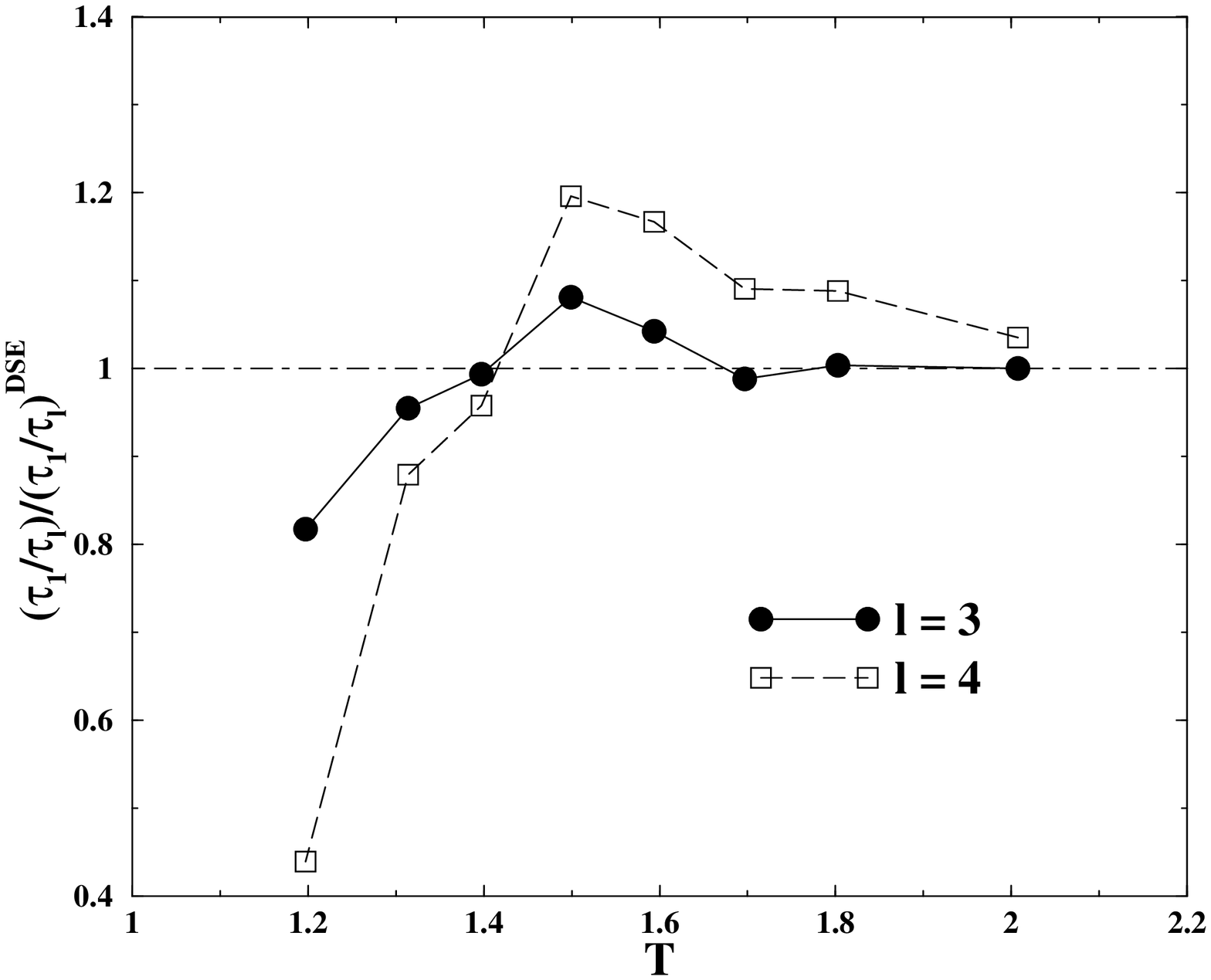}
\caption{\label{cap:tmp-tau3b1-tau4b1} The temperature dependence of the 
ratio between the first-rank and the $l$th rank orientational correlation 
times across the I-N transition for $l = 3$, and $4$. The inclusion of the 
scaling factor ensures that the ratio is equal to unity in the Debye 
limit as shown by the dot-dashed line.}
\end{figure}
\begin{figure}
\includegraphics[ width=7.0cm,angle=270]{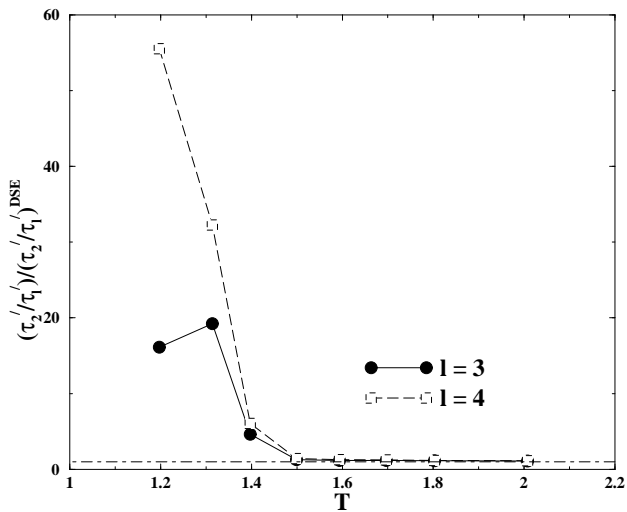}
\caption{\label{cap:tmp-tau3b2-tau4b2} The temperature dependence of 
the ratio between the second-rank and the $l$th rank orientational 
correlation times across the I-N transition for $l = 3$, and $4$. 
The inclusion of the scaling factor ensures that the ratio is equal to 
unity in the Debye limit as shown by the dot-dashed line.}
\end{figure}
In this work, we have also studied orientational relaxation in a system
of 500 Gay-Berne ellipsoids of revolution with the use of the same 
parameterization but along an isochore at the density $\rho=0.32$. The drop in 
temperature drives the system from the isotropic to the nematic phase with the
growth in the orientational order. In particular, the single-particle 
second-rank orientational correlation function decays with such a long time 
scale near the I-N phase boundary that an integral second-rank orientational 
correlation time is beyond the scope of the present simulation study. The poor 
data quality at long times with the present run length restricts us from having 
a reasonable fit of the long-time decay. In order to have an estimate of the 
second-rank orientational correlation time for the sake of comparison of the 
rank-dependent orientational correlation times, we define a correlation time 
$\tau_{l}^{/}(T)$ that is the time taken for the single-particle $l-$th rank 
orientational correlation function to decay by 90\% at a temperate $T$. The 
prime is used to distinguish it from the corresponding integral orientational 
correlation time. 

Fig. \ref{cap:tmp-tau2b1-op} shows that the ratio 
$\tau_{1}^{/}/\tau_{2}^{/}$ follows the Debye behavior away from the I-N 
transition, but the onset of the rapid growth of the orientational order 
parameter near the I-N transition induces a marked deviation from the Debye 
limit and the ratio falls rapidly. On the other hand, the ratios 
$\tau_{1}/\tau_{l}$ go through maxima on transit from the isotropic phase 
to the nematic phase for both $l = 3$ and $l = 4$ as shown in 
Fig. \ref{cap:tmp-tau3b1-tau4b1}, and the maxima correspond to the 
temperature below which the orientational order parameter is on the rapid rise.
Figure \ref{cap:tmp-tau3b2-tau4b2} illustrates the temperature behavior of the 
ratios $\tau_{2}^{/}/\tau_{l}^{/}$ across the I-N transition for $l = 3$, 
and $4$. While only a small deviation from the Debye behavior is apparent even 
at high temperatures away from the I-N transition, the onset of the growth of 
the orientational parameter marks a {\it sharp increase} in these ratios. The 
results, embodied in Figs. \ref{cap:tmp-tau2b1-op}-\ref{cap:tmp-tau3b2-tau4b2}, suggest that orientational correlation, that builds up across the I-N 
transition, plays a key role in deviation from the Debye behavior of the 
orientational correlation times.

The contrast between the study along an isotherm and that along an isochore
reveals the importance of the role played by the intermolecular
potential in the breakdown of the DSE relation. In the study along the 
isotherm, the free volume that is available for the rotation reduces thus
leading to the formation of the orientational caging. In contrast, 
when temperature is reduced, the attractive part of the
inter molecular potential becomes more dominant and results in the formation
of the orientational caging that arrests the orientational random walk
of the molecules. The above difference arises because temperature 
variation leads only to small changes in density because of the dominance
of the repulsive part of the potential in determining the liquid structure.

\section{Theoretical analysis of orientation relaxation \label{sec:ts}}

Here we present a mode coupling theory (MCT) analysis of the above
relaxation behaviour. MCT has a long and honorable history in describing
dynamics near phase transitions \cite{mct:ma,tdcp,bb:acp2}. Our starting point
of the theoretical analysis is the Zwanzig-Mori continued
fraction representation of the frequency dependent orientational time
correlation function, $C_{l}(z)$ \cite{bb:ch1,bb:rv4,hab_vol,bb:tml},
\begin{equation}
C_{l}^{s}(z)=\frac{1}{z+\frac{l(l+1)k_{B}T}{I(z+\Gamma_{l}(z))}}
\label{eq:clz}
\end{equation}
where $I$ is the moment of inertia and $\Gamma_{l}(z)$ is the Laplace frequency
($z$) and rank dependent memory function. The latter is determined
by the torque-torque correlation function. In general, it is very difficult
to determine this correlation function from first principles, but
as a first approximation, we would combine input from the mode coupling theory 
with that from the time dependent density functional theory to obtain an expression
for the memory function, $\Gamma_{l}(z)$, which can be
used to understand the reasons for the breakdown of the DSE model. Near
the I-N transition the memory function 
$\Gamma(z)$ 
can be written as a sum of two parts 
\begin{equation}
\Gamma_{l}(z)=\Gamma^{bare}+\Gamma_{l}^{sing}(z),
\label{eq:Gamma}
\end{equation}
where the bare part of the memory function is assumed to be rank
and frequency independent. This can be described by the two-body (binary) collision
model. Note that in conjunction with Eq.\ref{eq:clz}, $\Gamma^{bare}$ leads
to the DSE behaviour. The singular part of the memory function contains
effects of intermolecular correlation and is rank dependent. It is
given by \cite{bb:tml,bb:fay1,bb:fay2},
\begin{equation}
\Gamma_{l}^{sing}(z)=\frac{3k_BT\rho}{8\pi^3I}\int_{0}^{\infty}dt\;
e^{-zt}\int_{0}^{\infty}dk\; k^{2}\sum_{m}c_{llm}^{2}(k)
F_{lm}(k,t).
\label{eq:sing}
\end{equation}
In the above equation,  $\Gamma_l$ is the rank and frequency dependent
memory function. This is a function of the $l,l,m$ component of the wavevector
dependent direct correlation function $c_{llm}(k)$ (in the inter
molecular frame). $F_{lm}(k,t)$ is the $l,m$ component of the orientation dependent 
self-intermediate scattering function. The $F_{lm}(k,t)$ is
defined in terms of the spherical harmonics as 
\begin{equation}
F_{lm}(k,t)=
\left<e^{{\bf k}\cdot({\bf r(t)-{\bf r(0)}})}Y_{lm}(\Omega,0)Y_{lm}(\Omega,t)\right>.
\label{eq:fllm}
\end{equation}
The single particle position and orientation dependent $\Gamma({\bf r}-{\bf r^{\prime}},t-t^{\prime},{\bf \Omega},{\bf \Omega^{\prime}})$
memory function is related to the torque-torque correlation function through
the density functional theory by following the fluctuation dissipation
theorem \cite{bb:ch1,bb:tml}.
Note that $\Gamma_{l}^{sing}(z)$
contains the integration over the wave vector dependence. The slow down
of the relaxation of single particle orientational correlation function
is related to the nature of the component of the dynamics structure factor.

It is important to note that the rotational friction depends on the rank
of the orientational correlation function and this friction
differs from rank to rank because of vastly different
wave vector dependence of $F_{\ell m}(k,t)$, for different $\ell$,
particularly at low wavenumbers. Near the I-N transition,
due to existence of large wave length fluctuations, the $k\rightarrow0$
component of the $F_{20}(k,t)$ undergoes a very slow decay and 
this is responsible for the slow
down of the relaxation of the $C^{s}_{2}(t)$. 
In this limit, the expression for $F_{lm}^{s}(k,t)$
is given by \cite{bb:ch1}\begin{equation}
F_{lm}(k,t)=S_{lm}(k)e^{-\frac{l(l+1)D_{R}t}{S_{lm}(k)}},
\label{eq:fs22}
\end{equation}
where $S_{lm}(k)$ is the orientation dependent structure
factor and $D_{R}$ is the rotational diffusion coefficient. The $S_{lm}(k)$
is given by the expression
\begin{equation}
S_{lm}(k)=\left<e^{{\bf k}\cdot({\bf r-{\bf r^{\prime}}})}Y_{lm}(\Omega,0)Y_{lm}
(\Omega^{\prime},t)\right>
\label{eq:sllmk}
\end{equation}
Near the I-N transition, $S_{20}(k)$ grows as $1/B^{2}k^{2}$ ($B=\frac{\rho}{4\pi}\left(\frac{d^{2}c_{ll^{\prime}m}(k)}{dk^{2}}\right)_{k=0}$).
\begin{figure}
\includegraphics[width=8.3cm]{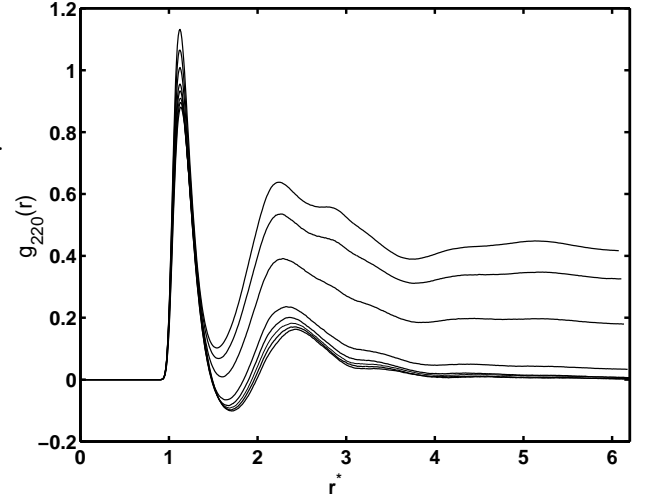}
\caption{\label{cap:g220} The $g_{220}$ component of the pair correlation 
function versus the pair separation at several densities across the I-N 
transition. The curves staring from the bottom to the top are corresponding
to densities between $\rho^{*}=0.285$ and $\rho^{*}=0.315$ on a grid of 
$\delta\rho^{*}=0.005$.}
\end{figure}
 The growth 
of orientational pair correlation with the approach of the I-N transition is 
evident in figure \ref{cap:g220}. Note that the starting from density 
$\rho^{*}$=0.305, the orientational pair correlation function 
becomes non-zero even at large intermolecular separations. This
is reflected in the rapid growth of $S_{20}(k)$ as $k \rightarrow 0$
near the I-N transition.

 Combination of the above factors provides the following simple 
expression for the frequency dependence of the singular part of the memory kernel
\begin{equation}
\Gamma^{sing}_{2}(z)=A/\sqrt{z}
\end{equation}
where A is a numerical constant \cite{bb:fay1}. Note that such a power
law dependence is absent from all odd $\ell$-th $\Gamma_{\ell}$ but in
principle present in all the even $\ell$. However, as $\ell$
increases, the decay becomes increasingly inertial and the role
of intermolecular correlation becomes weak beyond $\ell =4$. But for
$\ell=2$ and $\ell=4$, the inverse square root dependence of the
rotational memory function leads to a markedly slower, power law decay,
as seen from figure 2(b). 

 Physically, this power law is a manifestation
of the growth of the pseudo-nematic domains near the I-N phase
boundary. A particle inside
this domain feels a localizing potential which makes its
rotation difficult. However, even within such a domain, a rotation
of an individual particle by 180\textdegree (that is, by $\pi$) is possible
because of the up-down symmetry of the particle. However, in contradiction
to the conceptualization prevalent in supercooled liquid literature,
here such a half-cycle rotation relaxes only the odd rank correlation functions, leaving the even ranked ones unchanged. In the MCT description,
the influence of the localizing
potential enters through the two particle direct correlation function 
($c_{\ell \ell m}(k)$) and the static structure factor $S_{\ell m}(k)$.
Just as in the Maier-Saupe theory, the present mean-field theory description
can capture the {\it rank dependence of the effective, localizing 
potential.}
 
The above mode coupling theory analysis is by no means complete, but
it provides a semi-quantitative explanation of the observed 
rank dependence of
the orientational correlation time near the I-N transition, 
in terms of the rapid growth of equilibrium orientational 
pair correlation function.

\section{Concluding remarks\label{sec:cr}}

Let us first summarize the main results presented here. We study a system of 
Gay-Berne model mesogens along an isotherm and an isochore separately across
its isotropic-nematic phase transition to investigate the rank dependent 
single-particle orientational relaxation from the perspective of the Debye
behavior. Our results demonstrate that orientational correlation that starts
growing near the I-N transition as it is approached from the isotropic side
induces a marked deviation from the Debye behavior. We present a theoretical
analysis of our results within the framework of the mode-coupling theory. This 
mode coupling theory analysis provides a semi-quantitative explanation of the 
observed rank dependence of the orientational correlation time near the I-N 
transition. This explanation does not invoke the existence of any large scale
angular jump motion which randomizes and thereby leads to the decay 
of correlation  of all ranks with the same rate. Instead, our analysis
attributes the non-Debye rank dependence to the rapid growth of equilibrium 
orientational pair correlation function. 

As already mentioned, an earlier onset of stretching in the even rank 
correlation functions than in the corresponding odd rank functions has been 
observed in supercooled liquids as well \cite{scool:kam,scool:lap2}. This 
similarity is indeed interesting and deserves further study.

In view of the results presented here, a comparative study between 
dielectric relaxation (which essentially measures the $\ell =1$ correlation
function) and light scattering or fluorescence depolarization or even
NMR (all of these measure the $\ell =2$ correlation function) should
be worthwhile. In fact, a detailed theoretical analysis of dielectric
relaxation near the I-N phase boundary seems to be lacking.

\begin{acknowledgments}

This work was supported in parts by grants from DST and CSIR, India. DC 
acknowledges the UGC, India for providing financial support. We thank Dr. 
Sarika M. Bhattacharyya for discussions.

\end{acknowledgments}


\begin{thebibliography}{40}
\expandafter\ifx\csname natexlab\endcsname\relax\def\natexlab#1{#1}\fi
\expandafter\ifx\csname bibnamefont\endcsname\relax
  \def\bibnamefont#1{#1}\fi
\expandafter\ifx\csname bibfnamefont\endcsname\relax
  \def\bibfnamefont#1{#1}\fi
\expandafter\ifx\csname citenamefont\endcsname\relax
  \def\citenamefont#1{#1}\fi
\expandafter\ifx\csname url\endcsname\relax
  \def\url#1{\texttt{#1}}\fi
\expandafter\ifx\csname urlprefix\endcsname\relax\def\urlprefix{URL }\fi
\providecommand{\bibinfo}[2]{#2}
\providecommand{\eprint}[2][]{\url{#2}}

\bibitem[{\citenamefont{Debye}(1929)}]{debye}
\bibinfo{author}{\bibfnamefont{P.}~\bibnamefont{Debye}},
  \emph{\bibinfo{title}{Polar Molecules}} (\bibinfo{publisher}{Dover
  Publications, INC.}, \bibinfo{address}{New York}, \bibinfo{year}{1929}).

\bibitem[{\citenamefont{Berne and Pecora}(1976)}]{bap}
\bibinfo{author}{\bibfnamefont{B.~J.} \bibnamefont{Berne}} \bibnamefont{and}
  \bibinfo{author}{\bibfnamefont{R.}~\bibnamefont{Pecora}},
  \emph{\bibinfo{title}{Dynamic Light Scattering: With applications to
  Chemistry, Biology and Physics.}} (\bibinfo{publisher}{John Wiley \& Sons,
  INC}, \bibinfo{address}{New York}, \bibinfo{year}{1976}).

\bibitem[{\citenamefont{Cannistraro and Sacchetti}(1976)}]{alb-dbye}
\bibinfo{author}{\bibfnamefont{S.}~\bibnamefont{Cannistraro}} \bibnamefont{and}
  \bibinfo{author}{\bibfnamefont{F.}~\bibnamefont{Sacchetti}},
  \bibinfo{journal}{Phys. Rev. A} \textbf{\bibinfo{volume}{33}},
  \bibinfo{pages}{745} (\bibinfo{year}{1976}).

\bibitem[{\citenamefont{Paparo et~al.}(2002)\citenamefont{Paparo, Manzo,
  Marrucci, and Kreuzer}}]{shape:debye}
\bibinfo{author}{\bibfnamefont{D.}~\bibnamefont{Paparo}},
  \bibinfo{author}{\bibfnamefont{C.}~\bibnamefont{Manzo}},
  \bibinfo{author}{\bibfnamefont{L.}~\bibnamefont{Marrucci}}, \bibnamefont{and}
  \bibinfo{author}{\bibfnamefont{M.}~\bibnamefont{Kreuzer}},
  \bibinfo{journal}{J. Chem. Phys.} \textbf{\bibinfo{volume}{117}},
  \bibinfo{pages}{2187} (\bibinfo{year}{2002}).

\bibitem[{\citenamefont{Zwanzig}(1978)}]{zwanzig:debye1}
\bibinfo{author}{\bibfnamefont{R.}~\bibnamefont{Zwanzig}}, \bibinfo{journal}{J.
  Chem. Phys.} \textbf{\bibinfo{volume}{68}}, \bibinfo{pages}{4325}
  (\bibinfo{year}{1978}).

\bibitem[{\citenamefont{Stillinger and Hodgdon}(1994)}]{still1:ng}
\bibinfo{author}{\bibfnamefont{F.~H.} \bibnamefont{Stillinger}}
  \bibnamefont{and} \bibinfo{author}{\bibfnamefont{J.~A.}
  \bibnamefont{Hodgdon}}, \bibinfo{journal}{Phys. Rev. E}
  \textbf{\bibinfo{volume}{50}}, \bibinfo{pages}{2064} (\bibinfo{year}{1994}).

\bibitem[{\citenamefont{Stillinger and Hodgdon}(1996)}]{still2:ng}
\bibinfo{author}{\bibfnamefont{F.~H.} \bibnamefont{Stillinger}}
  \bibnamefont{and} \bibinfo{author}{\bibfnamefont{J.~A.}
  \bibnamefont{Hodgdon}}, \bibinfo{journal}{Phys. Rev. E}
  \textbf{\bibinfo{volume}{53}}, \bibinfo{pages}{2995} (\bibinfo{year}{1996}).

\bibitem[{\citenamefont{Tarjus and Kivelson}(1995)}]{scool:kiv1}
\bibinfo{author}{\bibfnamefont{G.}~\bibnamefont{Tarjus}} \bibnamefont{and}
  \bibinfo{author}{\bibfnamefont{D.}~\bibnamefont{Kivelson}},
  \bibinfo{journal}{J. Chem. Phys.} \textbf{\bibinfo{volume}{103}},
  \bibinfo{pages}{3071} (\bibinfo{year}{1995}).

\bibitem[{\citenamefont{Kammerer et~al.}(1997)\citenamefont{Kammerer, Kob, and
  Schilling}}]{scool:kam}
\bibinfo{author}{\bibfnamefont{S.}~\bibnamefont{Kammerer}},
  \bibinfo{author}{\bibfnamefont{W.}~\bibnamefont{Kob}}, \bibnamefont{and}
  \bibinfo{author}{\bibfnamefont{R.}~\bibnamefont{Schilling}},
  \bibinfo{journal}{Phys. Rev. E} \textbf{\bibinfo{volume}{56}},
  \bibinfo{pages}{5450} (\bibinfo{year}{1997}).

\bibitem[{\citenamefont{Ngai}(1999)}]{scool:ngai1}
\bibinfo{author}{\bibfnamefont{K.~L.} \bibnamefont{Ngai}}, \bibinfo{journal}{J.
  Phys. Chem. B} \textbf{\bibinfo{volume}{103}}, \bibinfo{pages}{10684}
  (\bibinfo{year}{1999}).

\bibitem[{\citenamefont{Ediger}(2000)}]{scool:edig1}
\bibinfo{author}{\bibfnamefont{M.~D.} \bibnamefont{Ediger}},
  \bibinfo{journal}{Ann. Rev. Phys. Chem.} \textbf{\bibinfo{volume}{51}},
  \bibinfo{pages}{99} (\bibinfo{year}{2000}).

\bibitem[{\citenamefont{Sillescue}(1999)}]{scool:sill2}
\bibinfo{author}{\bibfnamefont{H.}~\bibnamefont{Sillescue}},
  \bibinfo{journal}{J. Non. Crst. Sol.} \textbf{\bibinfo{volume}{243}},
  \bibinfo{pages}{81} (\bibinfo{year}{1999}).

\bibitem[{\citenamefont{Michele and Leporini}(2001)}]{scool:lap2}
\bibinfo{author}{\bibfnamefont{C.~D.} \bibnamefont{Michele}} \bibnamefont{and}
  \bibinfo{author}{\bibfnamefont{D.}~\bibnamefont{Leporini}},
  \bibinfo{journal}{Phys. Rev. E} \textbf{\bibinfo{volume}{63}},
  \bibinfo{pages}{36702} (\bibinfo{year}{2001}).

\bibitem[{\citenamefont{Andereozzi et~al.}(1997)\citenamefont{Andereozzi,
  di~Schino, Giordano, and Leporini}}]{scool:lap3}
\bibinfo{author}{\bibfnamefont{L.}~\bibnamefont{Andereozzi}},
  \bibinfo{author}{\bibfnamefont{A.}~\bibnamefont{di~Schino}},
  \bibinfo{author}{\bibfnamefont{M.}~\bibnamefont{Giordano}}, \bibnamefont{and}
  \bibinfo{author}{\bibfnamefont{D.}~\bibnamefont{Leporini}},
  \bibinfo{journal}{Europhys. Lett.} \textbf{\bibinfo{volume}{38}},
  \bibinfo{pages}{669} (\bibinfo{year}{1997}).

\bibitem[{\citenamefont{Andreozziyz et~al.}(1996)\citenamefont{Andreozziyz,
  Schinoy, Giordanoyz, and Leporini}}]{scool:lap4}
\bibinfo{author}{\bibfnamefont{L.}~\bibnamefont{Andreozziyz}},
  \bibinfo{author}{\bibfnamefont{A.~D.} \bibnamefont{Schinoy}},
  \bibinfo{author}{\bibfnamefont{M.}~\bibnamefont{Giordanoyz}},
  \bibnamefont{and} \bibinfo{author}{\bibfnamefont{D.}~\bibnamefont{Leporini}},
  \bibinfo{journal}{J. Phys.: Condens. Matter} \textbf{\bibinfo{volume}{8}},
  \bibinfo{pages}{9605} (\bibinfo{year}{1996}).

\bibitem[{\citenamefont{Bielowka et~al.}(2001)\citenamefont{Bielowka, Psurek,
  Ziolo, and Paluch}}]{scool:fdebye}
\bibinfo{author}{\bibfnamefont{S.~H.} \bibnamefont{Bielowka}},
  \bibinfo{author}{\bibfnamefont{T.}~\bibnamefont{Psurek}},
  \bibinfo{author}{\bibfnamefont{J.}~\bibnamefont{Ziolo}}, \bibnamefont{and}
  \bibinfo{author}{\bibfnamefont{M.}~\bibnamefont{Paluch}},
  \bibinfo{journal}{Phys. Rev. E} \textbf{\bibinfo{volume}{63}},
  \bibinfo{pages}{062301} (\bibinfo{year}{2001}).

\bibitem[{\citenamefont{Corezzi et~al.}(1999)\citenamefont{Corezzi, Capaccioli,
  Gallone, Lucchesi, and Rolla}}]{scool:corez}
\bibinfo{author}{\bibfnamefont{S.}~\bibnamefont{Corezzi}},
  \bibinfo{author}{\bibfnamefont{S.}~\bibnamefont{Capaccioli}},
  \bibinfo{author}{\bibfnamefont{G.}~\bibnamefont{Gallone}},
  \bibinfo{author}{\bibfnamefont{M.}~\bibnamefont{Lucchesi}}, \bibnamefont{and}
  \bibinfo{author}{\bibfnamefont{P.~A.} \bibnamefont{Rolla}},
  \bibinfo{journal}{J. Phys.: Condens. Matter} \textbf{\bibinfo{volume}{11}},
  \bibinfo{pages}{10297} (\bibinfo{year}{1999}).

\bibitem[{\citenamefont{Gottke et~al.}(2002{\natexlab{a}})\citenamefont{Gottke,
  Brace, Cang, Bagchi, and Fayer}}]{bb:fay1}
\bibinfo{author}{\bibfnamefont{S.~D.} \bibnamefont{Gottke}},
  \bibinfo{author}{\bibfnamefont{D.~D.} \bibnamefont{Brace}},
  \bibinfo{author}{\bibfnamefont{H.}~\bibnamefont{Cang}},
  \bibinfo{author}{\bibfnamefont{B.}~\bibnamefont{Bagchi}}, \bibnamefont{and}
  \bibinfo{author}{\bibfnamefont{M.~D.} \bibnamefont{Fayer}},
  \bibinfo{journal}{J. Chem. Phys.} \textbf{\bibinfo{volume}{116}},
  \bibinfo{pages}{360} (\bibinfo{year}{2002}{\natexlab{a}}).

\bibitem[{\citenamefont{Gottke et~al.}(2002{\natexlab{b}})\citenamefont{Gottke,
  Cang, Bagchi, and Fayer}}]{bb:fay2}
\bibinfo{author}{\bibfnamefont{S.~D.} \bibnamefont{Gottke}},
  \bibinfo{author}{\bibfnamefont{H.}~\bibnamefont{Cang}},
  \bibinfo{author}{\bibfnamefont{B.}~\bibnamefont{Bagchi}}, \bibnamefont{and}
  \bibinfo{author}{\bibfnamefont{M.~D.} \bibnamefont{Fayer}},
  \bibinfo{journal}{J. Chem. Phys.} \textbf{\bibinfo{volume}{116}},
  \bibinfo{pages}{6339} (\bibinfo{year}{2002}{\natexlab{b}}).

\bibitem[{\citenamefont{Cang et~al.}(2002)\citenamefont{Cang, Li, and
  Fayer}}]{fay2}
\bibinfo{author}{\bibfnamefont{H.}~\bibnamefont{Cang}},
  \bibinfo{author}{\bibfnamefont{J.}~\bibnamefont{Li}}, \bibnamefont{and}
  \bibinfo{author}{\bibfnamefont{M.~D.} \bibnamefont{Fayer}},
  \bibinfo{journal}{Chem. Phys. Lett.} \textbf{\bibinfo{volume}{366}},
  \bibinfo{pages}{82} (\bibinfo{year}{2002}).

\bibitem[{\citenamefont{Cang et~al.}(2003)\citenamefont{Cang, Li, Novikov, and
  Fayer}}]{fay3}
\bibinfo{author}{\bibfnamefont{H.}~\bibnamefont{Cang}},
  \bibinfo{author}{\bibfnamefont{J.}~\bibnamefont{Li}},
  \bibinfo{author}{\bibfnamefont{V.~N.} \bibnamefont{Novikov}},
  \bibnamefont{and} \bibinfo{author}{\bibfnamefont{M.~D.} \bibnamefont{Fayer}},
  \bibinfo{journal}{J. Chem. Phys.} \textbf{\bibinfo{volume}{118}},
  \bibinfo{pages}{9303} (\bibinfo{year}{2003}).

\bibitem[{\citenamefont{Jose and Bagchi}(2004{\natexlab{a}})}]{bb:ppj1}
\bibinfo{author}{\bibfnamefont{P.~P.} \bibnamefont{Jose}} \bibnamefont{and}
  \bibinfo{author}{\bibfnamefont{B.}~\bibnamefont{Bagchi}},
  \bibinfo{journal}{J. Chem. Phys.} \textbf{\bibinfo{volume}{120}},
  \bibinfo{pages}{11256} (\bibinfo{year}{2004}{\natexlab{a}}).

\bibitem[{\citenamefont{Jose et~al.}(2005)\citenamefont{Jose, Chakrabarti, and
  Bagchi}}]{bb:ppj3}
\bibinfo{author}{\bibfnamefont{P.~P.} \bibnamefont{Jose}},
  \bibinfo{author}{\bibfnamefont{D.}~\bibnamefont{Chakrabarti}},
  \bibnamefont{and} \bibinfo{author}{\bibfnamefont{B.}~\bibnamefont{Bagchi}},
  \bibinfo{journal}{Phys. Rev. E} \textbf{\bibinfo{volume}{71}},
  \bibinfo{pages}{030701} (\bibinfo{year}{2005}).

\bibitem[{\citenamefont{Chakrabarti et~al.}(2005)\citenamefont{Chakrabarti,
  Jose, Chakrabarty, and Bagchi}}]{bb:dwc}
\bibinfo{author}{\bibfnamefont{D.}~\bibnamefont{Chakrabarti}},
  \bibinfo{author}{\bibfnamefont{P.~P.} \bibnamefont{Jose}},
  \bibinfo{author}{\bibfnamefont{S.}~\bibnamefont{Chakrabarty}},
  \bibnamefont{and} \bibinfo{author}{\bibfnamefont{B.}~\bibnamefont{Bagchi}},
  \bibinfo{journal}{Phys. Rev. Lett.} \textbf{\bibinfo{volume}{95}},
  \bibinfo{pages}{197801} (\bibinfo{year}{2005}).

\bibitem[{\citenamefont{DeMiguel et~al.}(1992)\citenamefont{DeMiguel, Rull, and
  Gubbins}}]{deMug2}
\bibinfo{author}{\bibfnamefont{E.}~\bibnamefont{DeMiguel}},
  \bibinfo{author}{\bibfnamefont{L.~F.} \bibnamefont{Rull}}, \bibnamefont{and}
  \bibinfo{author}{\bibfnamefont{K.~E.} \bibnamefont{Gubbins}},
  \bibinfo{journal}{Phys. Rev. A} \textbf{\bibinfo{volume}{45}},
  \bibinfo{pages}{3813} (\bibinfo{year}{1992}).

\bibitem[{\citenamefont{Vasanthi et~al.}(2001)\citenamefont{Vasanthi,
  Bhattacharyya, and Bagchi}}]{bb:vas2}
\bibinfo{author}{\bibfnamefont{R.}~\bibnamefont{Vasanthi}},
  \bibinfo{author}{\bibfnamefont{S.}~\bibnamefont{Bhattacharyya}},
  \bibnamefont{and} \bibinfo{author}{\bibfnamefont{B.}~\bibnamefont{Bagchi}},
  \bibinfo{journal}{J. Chem. Phys.} \textbf{\bibinfo{volume}{116}},
  \bibinfo{pages}{1092} (\bibinfo{year}{2001}).

\bibitem[{\citenamefont{Jose and Bagchi}(2004{\natexlab{b}})}]{bb:ppj2}
\bibinfo{author}{\bibfnamefont{P.~P.} \bibnamefont{Jose}} \bibnamefont{and}
  \bibinfo{author}{\bibfnamefont{B.}~\bibnamefont{Bagchi}},
  \bibinfo{journal}{J. Chem. Phys.} \textbf{\bibinfo{volume}{121}},
  \bibinfo{pages}{6978} (\bibinfo{year}{2004}{\natexlab{b}}).

\bibitem[{\citenamefont{Berne and Pechukas}(1972)}]{berne1}
\bibinfo{author}{\bibfnamefont{B.~J.} \bibnamefont{Berne}} \bibnamefont{and}
  \bibinfo{author}{\bibfnamefont{P.}~\bibnamefont{Pechukas}},
  \bibinfo{journal}{J. Chem. Phys.} \textbf{\bibinfo{volume}{56}},
  \bibinfo{pages}{4213} (\bibinfo{year}{1972}).

\bibitem[{\citenamefont{Gay and Berne}(1981)}]{berne2}
\bibinfo{author}{\bibfnamefont{J.~G.} \bibnamefont{Gay}} \bibnamefont{and}
  \bibinfo{author}{\bibfnamefont{B.~J.} \bibnamefont{Berne}},
  \bibinfo{journal}{J. Chem. Phys.} \textbf{\bibinfo{volume}{74}},
  \bibinfo{pages}{3316} (\bibinfo{year}{1981}).

\bibitem[{\citenamefont{de~Miguel and Vega}(2005)}]{deMug3}
\bibinfo{author}{\bibfnamefont{E.}~\bibnamefont{de~Miguel}} \bibnamefont{and}
  \bibinfo{author}{\bibfnamefont{C.}~\bibnamefont{Vega}}, \bibinfo{journal}{J.
  Chem. Phys.} \textbf{\bibinfo{volume}{177}}, \bibinfo{pages}{6313}
  (\bibinfo{year}{2005}).

\bibitem[{\citenamefont{de~Gennes and Prost}(1993)}]{deg:5}
\bibinfo{author}{\bibfnamefont{P.~G.} \bibnamefont{de~Gennes}}
  \bibnamefont{and} \bibinfo{author}{\bibfnamefont{J.}~\bibnamefont{Prost}},
  \emph{\bibinfo{title}{The Physics of Liquid Crystals}}
  (\bibinfo{publisher}{Clarendon Press}, \bibinfo{address}{Oxford},
  \bibinfo{year}{1993}).

\bibitem[{\citenamefont{Chandrasekhar}(1977)}]{sch}
\bibinfo{author}{\bibfnamefont{S.}~\bibnamefont{Chandrasekhar}},
  \emph{\bibinfo{title}{Liquid Crystals}} (\bibinfo{publisher}{Cambridge
  University Press}, \bibinfo{address}{Cambridge}, \bibinfo{year}{1977}).

\bibitem[{\citenamefont{Stephen and Straley}(1974)}]{rmp1}
\bibinfo{author}{\bibfnamefont{M.~J.} \bibnamefont{Stephen}} \bibnamefont{and}
  \bibinfo{author}{\bibfnamefont{J.~P.} \bibnamefont{Straley}},
  \bibinfo{journal}{Rev. Mod. Phys.} \textbf{\bibinfo{volume}{46}},
  \bibinfo{pages}{617} (\bibinfo{year}{1974}).

\bibitem[{\citenamefont{k.~Ma and Mazenko}(1975)}]{mct:ma}
\bibinfo{author}{\bibfnamefont{S.}~\bibnamefont{k.~Ma}} \bibnamefont{and}
  \bibinfo{author}{\bibfnamefont{G.~F.} \bibnamefont{Mazenko}},
  \bibinfo{journal}{Phys. Rev. B} \textbf{\bibinfo{volume}{11}},
  \bibinfo{pages}{4077} (\bibinfo{year}{1975}).

\bibitem[{\citenamefont{Hohenberg and Halperin}(1977)}]{tdcp}
\bibinfo{author}{\bibfnamefont{P.~C.} \bibnamefont{Hohenberg}}
  \bibnamefont{and} \bibinfo{author}{\bibfnamefont{B.~I.}
  \bibnamefont{Halperin}}, \bibinfo{journal}{Rev. Mod. Phys.}
  \textbf{\bibinfo{volume}{49}}, \bibinfo{pages}{435} (\bibinfo{year}{1977}).

\bibitem[{\citenamefont{Bagchi and Bhattacharyya}(2001)}]{bb:acp2}
\bibinfo{author}{\bibfnamefont{B.}~\bibnamefont{Bagchi}} \bibnamefont{and}
  \bibinfo{author}{\bibfnamefont{S.}~\bibnamefont{Bhattacharyya}},
  \bibinfo{journal}{Adv. Chem. Phys.} \textbf{\bibinfo{volume}{116}},
  \bibinfo{pages}{67} (\bibinfo{year}{2001}).

\bibitem[{\citenamefont{Bagchi and Chandra}(1991)}]{bb:ch1}
\bibinfo{author}{\bibfnamefont{B.}~\bibnamefont{Bagchi}} \bibnamefont{and}
  \bibinfo{author}{\bibfnamefont{A.}~\bibnamefont{Chandra}},
  \bibinfo{journal}{Adv. Chem. Phys.} \textbf{\bibinfo{volume}{80}},
  \bibinfo{pages}{1} (\bibinfo{year}{1991}).

\bibitem[{\citenamefont{Ravichandran and Bagchi}(1995)}]{bb:rv4}
\bibinfo{author}{\bibfnamefont{S.}~\bibnamefont{Ravichandran}}
  \bibnamefont{and} \bibinfo{author}{\bibfnamefont{B.}~\bibnamefont{Bagchi}},
  \bibinfo{journal}{Int. Rev. Phys. Chem.} \textbf{\bibinfo{volume}{14}},
  \bibinfo{pages}{271} (\bibinfo{year}{1995}).

\bibitem[{\citenamefont{Hubbard and Wolynes}(1978)}]{hab_vol}
\bibinfo{author}{\bibfnamefont{J.~B.} \bibnamefont{Hubbard}} \bibnamefont{and}
  \bibinfo{author}{\bibfnamefont{P.~G.} \bibnamefont{Wolynes}},
  \bibinfo{journal}{J. Chem. Phys.} \textbf{\bibinfo{volume}{69}},
  \bibinfo{pages}{998} (\bibinfo{year}{1978}).

\bibitem[{\citenamefont{Bagchi}(1998)}]{bb:tml}
\bibinfo{author}{\bibfnamefont{B.}~\bibnamefont{Bagchi}}, \bibinfo{journal}{J.
  Mol. Liq.} \textbf{\bibinfo{volume}{77}}, \bibinfo{pages}{177}
  (\bibinfo{year}{1998}).

\end{thebibliography}

\end{document}